\documentclass[a4paper,preprintnumbers,floatfix,superscriptaddress,pra,twocolumn,showpacs,nofootinbib]{revtex4-1}
\usepackage{amsmath, amsthm, amssymb}
\usepackage{graphicx}
\usepackage{dcolumn}
\usepackage{bm}
\usepackage{color}
\usepackage{amsmath}
\usepackage{amsfonts}
\usepackage{amssymb}
\usepackage{bbm}
\usepackage{hyperref}
\usepackage[usenames,dvipsnames]{xcolor}

\newcommand{\eqnref}[1]{(\ref{#1})}
\newcommand{\figref}[1]{Fig.~\ref{#1}}

\newcommand{\vacr}{|vac\rangle}
\newcommand{\vacl}{\langle vac|}
\newcommand{\bra}[1]{\langle #1|}
\newcommand{\ket}[1]{|#1\rangle}

\newcommand{\ba}{\begin{eqnarray}}
\newcommand{\be}{\begin{equation}}
\newcommand{\ee}{\end{equation}}
\newcommand{\ea}{\end{eqnarray}}
\newcommand{\ban}{\begin{eqnarray*}}
\newcommand{\ean}{\end{eqnarray*}}

{\begin{framed}\begin{small}}
{\end{small}\end{framed}}

\DeclareGraphicsExtensions{.pdf,.png,.jpg}

\makeindex

\begin{document}

\title{Testing nonlocality of a single photon without a shared reference frame}
\date{\today}

\author{Jonatan Bohr Brask}\affiliation{ICFO-Institut de Ciencies Fotoniques, Av.~Carl Friedrich Gauss, 3, 08860 Castelldefels (Barcelona), Spain}
\author{Rafael Chaves}\affiliation{Institute for Physics, University of Freiburg, Rheinstrasse 10, D-79104 Freiburg, Germany}
\author{Nicolas Brunner}
\affiliation{D\'epartement de Physique Th\'eorique, Universit\'e de Gen\`eve, 1211 Gen\`eve, Switzerland}
\affiliation{H.H. Wills Physics Laboratory, University of Bristol, Bristol, BS8 1TL, United Kingdom}

\begin{abstract}
The question of testing the nonlocality of a single photon has raised much debate over the last years. The controversy is intimately related to the issue of providing a common reference frame for the observers to perform their local measurements. Here we address this point by presenting a simple scheme for demonstrating the nonlocality of a single photon which does not require a shared reference frame.
Specifically, Bell inequality violation can be obtained with certainty with unaligned devices, even if the relative frame fluctuates between each experimental run of the Bell test. Our scheme appears feasible with current technology, and may simplify the realization of quantum communication protocols based on single-photon entanglement.
\end{abstract}


\maketitle

\section{Introduction}

Entanglement is usually considered as a property of systems composed of multiple quantum particles. It appears however also in a conceptually different form that involves only a single quantum particle. For instance, when sending a single photon on a balanced beam-splitter, one gets the output state
\ba
\label{eq.single}
\ket{\psi}_{AB} = \frac{1}{\sqrt{2}}( \ket{0}_A  \ket{1}_B +  \ket{1}_A  \ket{0}_B ) ,
\ea
where the output modes of the beam-splitter are denoted $A$ and $B$, and $\ket{j}_{A,B}$ designates a state of $j$ photons in mode $A$, $B$ respectively.
Although initially controversial (see e.g.~\cite{vanEnk2005} for a good summary of this discussion)
single-photon entanglement now attracts much attention, in particular for applications in quantum communication \cite{Morin2012,Sangouard2012,Sangouard2011rev}.

A question of fundamental interest is whether a single photon (or more generally a single particle) can exhibit nonlocality. In other words, can the state \eqref{eq.single} violate a Bell inequality? Note that entanglement is necessary for Bell nonlocality but not sufficient in general \cite{Werner1989}, hence the question. This issue was first investigated by Tan, Walls and Collett \cite{Tan1991}, who proposed a scheme for refuting a certain class of local models. Subsequently, Hardy \cite{Hardy1994} proposed a better scheme, able to test the most general local model. However, a long debate followed \cite{GHZ1994reply,*Vaidman1994reply,*Hardy1994reply}, the main point of which being that Hardy's scheme involved additional particles for performing the local measurements of the Bell test (e.g. laser beams for performing homodyne measurements). In fact, these additional particles provide the common reference frame (between the distant observers) and should therefore be synchronized in phase, i.e. they must be coherent. The problem is that
the concept of optical coherence has been questioned, in particular by M\o lmer \cite{Molmer1997}, who suggested that coherence between optical beams should in fact be accounted for by entanglement between these beams---for an enlightening perspective on this debate, and its connection to reference frames, see \cite{Bartlett2006}. In the context of Hardy's scheme, the point of controversy is thus whether the observed nonlocal correlations are mediated by the state \eqref{eq.single} or by the additional particles providing the reference frame, as in practice the local oscillators originate from a common laser \cite{Hessmo2004,*Babichev2004,*Angelo2006}.

In order to address this issue (or rather to circumvent it), one may ask whether the nonlocality of single photon can be demonstrated without having a common reference frame. This question was first discussed by Dunningham and Vedral \cite{Dunningham2007}. Their scheme demonstrates single-photon entanglement, however, it fails at demonstrating Bell inequality violation \footnote{Note that the state of eq.~(7) in \cite{Dunningham2007}, after averaging over the unknown phase, is of the form $\rho = \frac{2}{3} \ket{\psi}\bra{\psi} + \frac{1}{3} \ket{00}\bra{00}$. This state does not violate the CHSH inequality, hence does not exhibit Hardy's nonlocality, as can be checked via the criteria of the Horodeckis \cite{Horodecki2009}. To the best of our knowledge, the state $\rho$ is not known to violate any Bell inequality.}.

Here we present a simple scheme for demonstrating the nonlocality of a single photon without the requirement of a shared reference frame, hence settling the debate. In our scheme, which builds upon recent work \cite{Brask2012}, the local measurements are implemented by combining optical displacements \cite{Banaszek1999} and single-photon detection. Bell inequality violation can be guaranteed even though the relative phase between the local displacement beams is not synchronized. In fact, it turns out that even when the relative phase fluctuates from run to run during the experiment, Bell inequality violation can be obtained with certainty, provided that the phase follows any non-uniform Gaussian distribution. That is, nonlocality can be demonstrated for arbitrarily strong Gaussian phase fluctuations. In addition, the scheme can display nonlocality for arbitrarily small displacements, i.e.~when essentially no additional particles are introduced in the experiment. Altogether this also shows that the nonlocality of a single photon is actually far more generic and robust than previously thought.

Beyond the fundamental interest, our work may also be relevant for applications. As mentioned above, single-photon entanglement is important in quantum information protocols, e.g.~quantum repeaters \cite{Sangouard2011rev}. However, witnessing and characterising it in practice is challenging. This is partly due to the complexity of aligning a common reference frame, a central issue for experimental quantum communications which is often overlooked in theoretical works. Our scheme shows how single-photon entanglement can be tested in a device-independent way, that is without placing assumptions on the functioning of the devices used in the protocol, and without a common reference frame, and as we argue below, could be feasible with current technology. Hence we believe it opens interesting perspectives for applications of single-photon entanglement.

\section{Scheme}

We consider a single photon split between $N$ spatial modes, i.e. in the state
\begin{align}
\label{eq.W}
\ket{W_N} = \frac{1}{\sqrt{N}}\big( & \ket{0,0,\ldots,1} + \ldots \nonumber \\
& + \ket{0,1,0,\ldots,0} + \ket{1,0,\ldots,0} \big) .
\end{align}
Such states have been produced experimentally using a heralded single photon source based on spontaneous down conversion and beam-splitters \cite{Choi2010}. Note that for two observers \eqnref{eq.W} reduces to the state \eqnref{eq.single}.

For testing the nonlocality of state \eqref{eq.W}, we use the Bell inequalities introduced by Werner-Wolf-Weinfurter-Zukowski-Brukner (W${}^3$ZB) \cite{Werner2001,*Weinfurter2001,*Zukowski2002}. These inequalities apply to a scenario involving $N$ observers, each having the choice between two possible dichotomic measurements (with outcomes $\pm1$), denoted $M_0^{(k)}$, $M_1^{(k)}$ for party $k$. For this scenario, all relevant full-correlation (i.e. featuring only $N$-party correlation terms) Bell inequalities can be compactly expressed in a single (nonlinear) inequality
\begin{equation}
\label{eq.WWWZB}
S =  2^{-N} \sum_{r} \left\vert  \sum_{{\bf s}} (-1)^{{\bf r}\cdot{\bf s}} \xi({\bf s}) \right\vert \leq 1 ,
\end{equation}
where ${\bf r}$ and ${\bf s}$ are vectors in $\{0,1\}^N$ and $\xi({\bf s}) = \langle M^{(1)}_{s_1} \cdots M^{(N)}_{s_N} \rangle$ the corresponding full-correlation function. Note that for $N=2$, the above inequality is equivalent to the well known Clauser-Horne-Shimony-Holt (CHSH) inequality.

For the measurements, we consider the scheme suggested in Ref.~\cite{Banaszek1999}, namely an optical displacement followed by (non-number-resolving) single-photon detection. In Refs.~\cite{Brask2012,Chaves2011} it was shown that \eqnref{eq.W} can violate W${}^3$ZB using this kind of measurements for any $N\geq 2$. Assigning $+1$/$-1$ to the no-click and click events respectively, the observable is given by $M_{D} = 2\ket{\alpha}\bra{\alpha} - 1$, where $\ket{\alpha}$ is a coherent state~\cite{Brask2012,Banaszek1999}. Physically, the displacements are implemented by mixing the signal with a coherent state from a local oscillator on a beam splitter with high transmission. Different measurement settings correspond to different choices for the amplitude and phase of the local oscillator, which determine the amplitude and phase of $\alpha = r e^{\imath\varphi}$. Restricting to the $0,1$-photon subspace, the observable can be put on matrix form
\begin{equation}
M_D\left(  r,\phi\right)  =%
\begin{pmatrix}
2e^{-r^{2}}-1 & 2e^{-r^{2}-\imath\varphi}r\\
2e^{-r^{2}+\imath\varphi}r & 2e^{-r^{2}}r^{2}-1
\end{pmatrix}
.
\end{equation}
It is instructive to compare the above observable with a standard projective qubit measurement, of the form
\begin{equation} \label{qubit}
M_P\left(  \theta,\phi\right)  =\frac{1}{2}%
\begin{pmatrix}
\cos\theta & e^{-\imath\varphi}\sin\theta\\
e^{\imath\varphi}\sin\theta & -\cos\theta
\end{pmatrix}
.
\end{equation}
One can readily see that these two measurements are equivalent up to second order in $\theta$, by replacing $r \rightarrow \theta/2$. In particular, displacement measurements can faithfully mimic projective qubit measurements close to the $z$ axis of the Bloch sphere. Hence (at least for $N=2$) our scheme can be viewed as the equivalent of the usual multi-particle Bell test. This indicates that that there is no fundamental difference between single particle and multi-particle entanglement.

A nontrivial but apriori necessary issue is the alignment of a common reference frame. In the case of projective qubit measurements, this amounts to aligning two angles (azimuthal and polar on the Bloch sphere), which requires exchange of information between the parties. For optical displacements the situation is different, since the phase and amplitude of the local oscillator play different roles. Indeed the amplitude of the optical beam is an absolute quantity which can be characterized locally by each observer and does not require any exchange of information between the parties. On the other hand, the phase of each local beam must be aligned with respect to some shared reference phase, which provides a common clock between the parties. Experimentally, this is achieved by distributing an intense reference oscillator to all observers \cite{Hessmo2004,*Babichev2004,*Angelo2006}. As we show below, the synchronization of the local oscillator can in fact be dispensed with, at the cost of adding one additional measurement setting for the first party.

\section{Guaranteed Bell violation without a shared frame}

For the sake of clarity, we start by discussing the case of two parties who employ a simple measurement strategy. The amplitudes of their measurement settings are given by 0 and $r$. Hence the observables of party $k$ are given by $M_0^{(k)} = M_D(0,0)$ and $M_1^{(k)} = M_D(r,\varphi_k)$. Note that for $r=0$ the measurements correspond to single-photon detection and the local oscillator phase is irrelevant (we set it to 0 for simplicity). The correlators are then given by
\begin{align}
\label{eq.corrs2part}
\xi(0,0) & = -1 , \nonumber \\
\xi(0,1) & =  \xi(1,0) = - e^{-r^2} \left( 1 - r^2 \right)  , \\
\xi(1,1) & = 1 - 2 e^{-r^2} ( 1+r^2 ) + 4 e^{-2 r^2} r^2 (1+\cos(\phi )) . \nonumber
\end{align}
Note that the first three correlators are phase independent, and the last one depends only on the relative phase $\phi = \varphi_1 -\varphi_2 $. Any phase acquired in transmission from the source to the parties can also be absorbed in $\phi$.

\begin{figure} [!t]
\centering
\includegraphics[width=\linewidth]{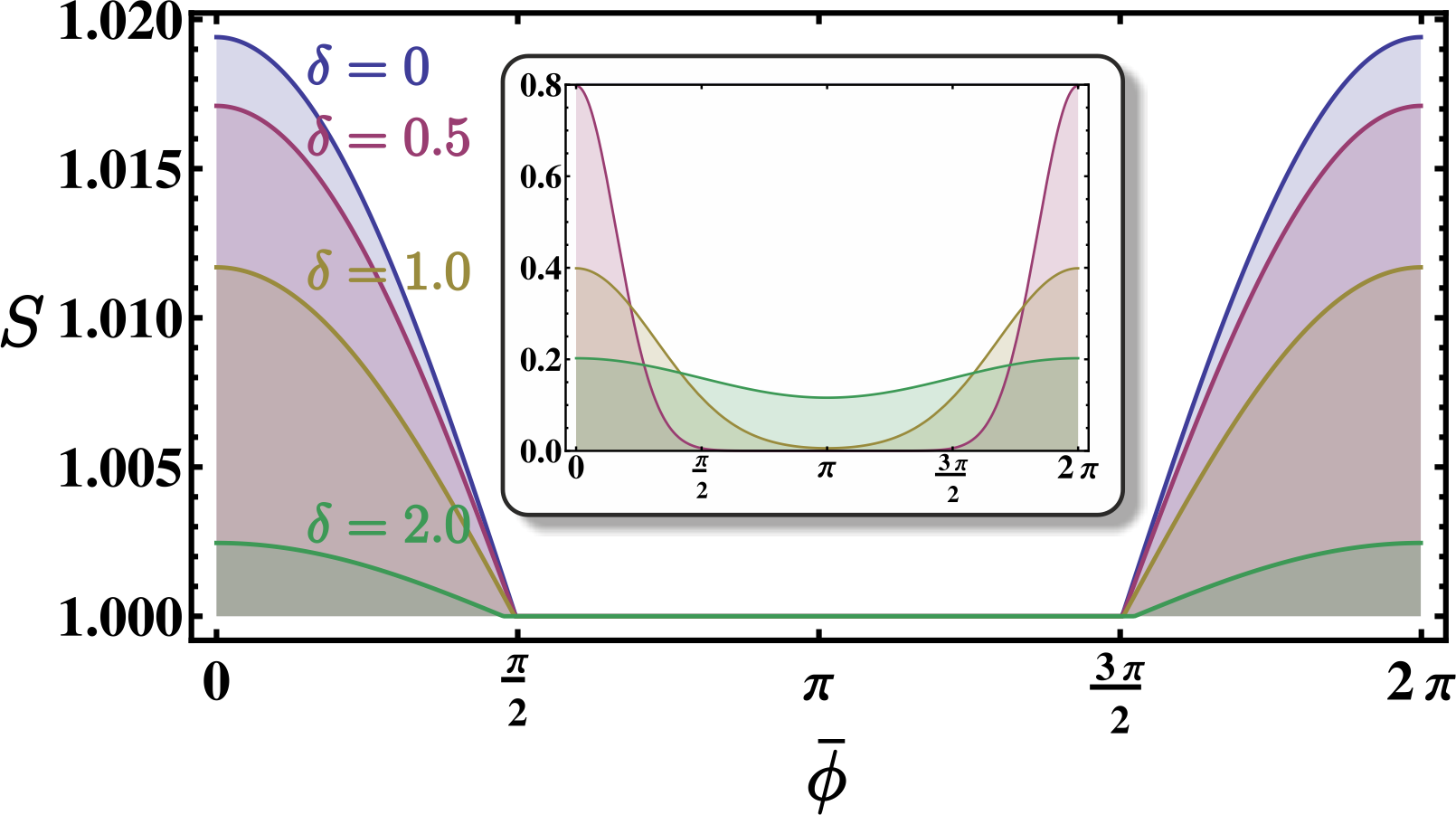}
\caption {(Color online) Bell value $S$ as a function of the average relative phase, for distributions of different width and $r=0.1$. The top curve corresponds to a static relative phase ($\delta=0$). For $\bar{\phi} \in ]0,\pi/2[ \cup ]3\pi/2,2\pi[$ the Bell inequality is violated, $S>1$. The inset shows the phase distributions, for $\bar{\phi} = 0$.}
\label{fig.smallrdistr}
\end{figure}

If the local oscillators have not been synchronized, the local phases will be independent. Let us first consider the situation in which the relative phase $\phi$ is uniformly distributed on the interval $[0,2\pi]$, but assumed to be fixed for the whole duration of the experiment, i.e.~it does not fluctuate from one run of the experiment to another. From the above, and for $r \ll 1$, we have
\begin{equation}
\label{eq.violsmallr}
S = 1 +  r^2   ( |\cos\phi| + \cos\phi ) .
\end{equation}
Hence, whenever $\phi \in ]0,\pi/2[ \cup ]3\pi/2,2\pi[$ we have a Bell inequality violation $S>1$ (see the blue curve in \figref{fig.smallrdistr}). We can ensure violation by introducing one additional setting for (say) the first party, measuring an observable with a shifted phase $M_2^{1} = M_D(r,\varphi_1+\pi)$ and then discarding either the runs where the first party employed $M_1^{1}$ or the runs with $M_2^{1}$. That is, for a static, unknown relative phase, nonlocality is detected with certainty, using only 3 settings for the first party and 2 settings for the second party.

Next, let us move to the case in which the relative phase fluctuates from run to run. We consider a Gaussian distribution of the phase with center $\bar{\phi}$ and width $\delta$. Since phases differing by an integer multiple of $2\pi$ are equivalent, the Gaussian must be wrapped onto the interval $\left[0,2\pi\right]$. The resulting distribution is $\frac{1}{2\pi} \vartheta( \frac{\phi-\bar{\phi}}{2} ; e^{-\delta^2/2} )$, where $\vartheta$ is the Jacobi theta function. A static relative phase corresponds to $\delta=0$ while a completely flat distribution is obtained for $\delta \rightarrow \infty$. Now, for a fluctuating phase the experimenters do not have access to the correlators  \eqnref{eq.corrs2part} but only to their averages over the phase distribution. To compute the average of $\xi(1,1)$, we use the fact that
\begin{equation}
\frac{1}{2\pi} \int_0^{2\pi} d\phi \, \vartheta( \frac{\phi-\bar{\phi}}{2} ; e^{-\delta^2/2} )  \cos\phi = e^{-\delta^2/2} \cos\bar{\phi} .
\end{equation}
For sufficiently small $r$, we then find that
\begin{equation}
\label{eq.violsmallrfluc}
S = 1 + e^{-\delta^2/2} r^2 ( |\cos \bar{\phi} | + \cos \bar{\phi} ) .
\end{equation}
From this expression we see that Bell violation is achieved for any gaussian distribution which is not completely flat ($\delta < \infty$), provided that $\bar{\phi} \in ]0,\pi/2[ \cup ]3\pi/2,2\pi[$ (see \figref{fig.smallrdistr}). That is, our scheme allows for violations even for arbitrary run-to-run fluctuations in the relative phase. As above, by adding one more measurement setting for the first party, we get Bell inequality violation with certainty, without any need for alignment. These results demonstrate a striking robustness of the nonlocality of single photon entanglement. Moreover, as there is here no coherence between the local oscillators, and since the latter may have an arbitrarily small amplitude, this unambiguously proves the nonlocal character of a single photon.

\begin{figure} [!t]
\centering
\includegraphics[width=\linewidth]{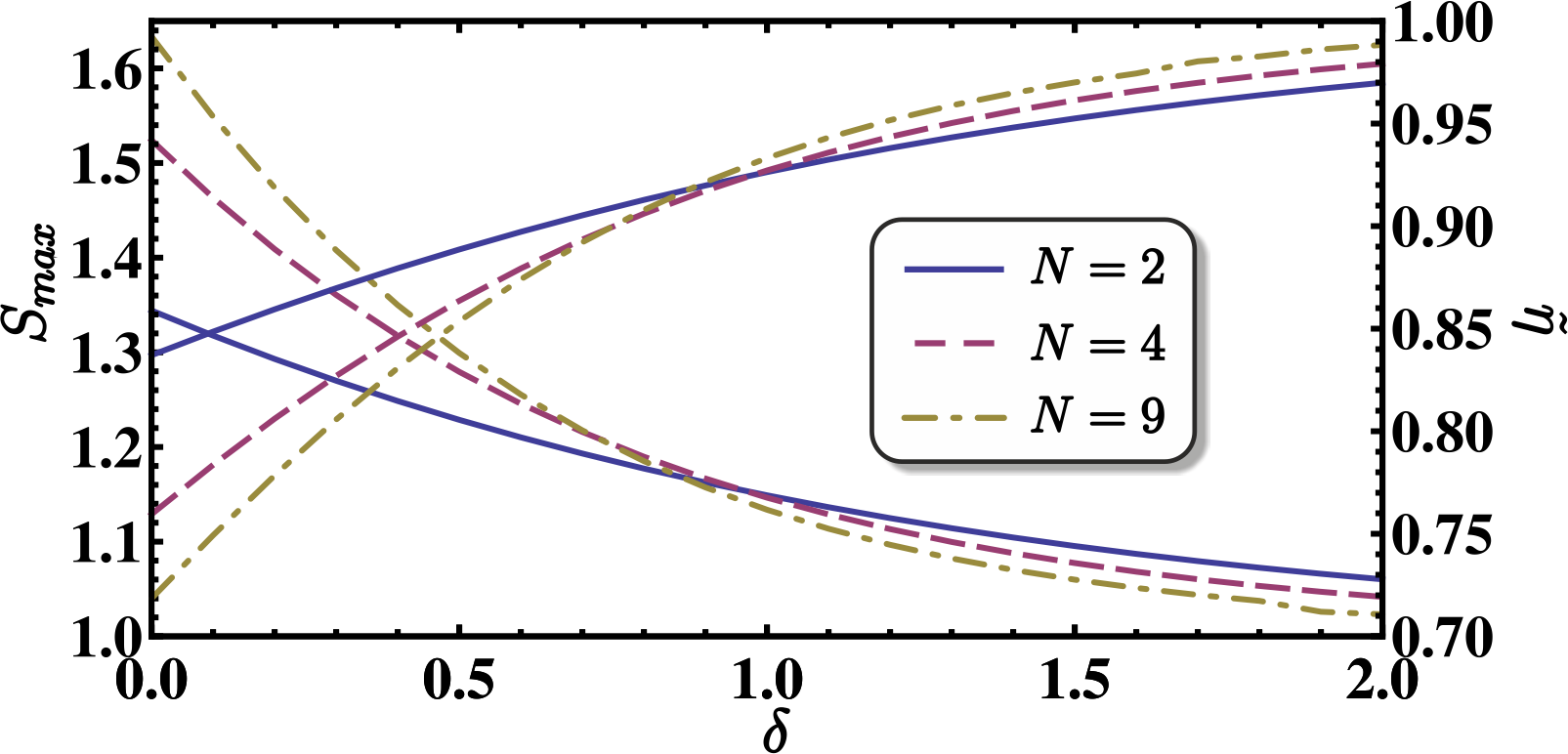}
\caption {(Color online) Maximal Bell value $S_ {max}$ (descending curves) and critical efficiency $\tilde{\eta}$ (ascending curves) vs.~phase fluctuations $\delta$ for different number of parties, $N=2,4,9$.
}
\label{fig.Smaxeta}
\end{figure}

\section{Practical perspective and more parties}

We discuss now the feasibility of our scheme, for the case of $N$ observers, taking into account the effect of losses. Similarly to above, we first compute the Bell violation $S$ for a given value of the average relative phase. Then, by adding extra measurement settings, we show how Bell violations can be obtained with certainty when the observers share no reference frame.

We allow all measurements to have non-zero intensity (i.e.~$r>0$), as this will lead to higher Bell violations. It turns out that it is always optimal for all parties to use the same amplitude settings, and for each party to use just a single phase. We thus compute $S$ for different numbers of parties, each party $k$ using measurements $M_0^{(k)} = M_D(r,\varphi_k)$ and $M_1^{(k)} = M_D(r',\varphi_k)$. The resulting expression depends on $N-1$ relative phases $\phi_l$. Assuming these relative phases to follow Gaussian distributions as above, with identical widths $\delta$ and centers $\bar{\phi}_l$, we find the maximal Bell value, $S_{max}$, for both static and fluctuating phases. The results are summarised in \figref{fig.Smaxeta} where the maximal violation is plotted against $\delta$. For the static case ($\delta=0$), the maximal violation for $N=2$ is $S \approx 1.34$, which is not far from the maximal quantum value of $S=\sqrt{2}$. Adding more parties, the Bell violation increases, provided the phase fluctuations are moderate ($\delta \lesssim 0.9$).

Next, we discuss the effect of losses, e.g.~in transmission or due to imperfect detectors. Assuming a total efficiency $\eta$ in each mode, the lossy state is of the form $\eta \ket{W_N}\bra{W_N} + (1- \eta) \vacr\vacl$, where $\vacr$ is the vacuum state. For given $N$ and $\delta$, we can numerically determine the threshold efficiency $\tilde{\eta}$ above which the Bell inequality W${}^3$ZB can be violated. Results are shown in \figref{fig.Smaxeta}. For moderate phase fluctuations ($\delta \lesssim 0.9$), increasing $N$ reduces the threshold efficiency. Note that for the present scheme Bell violations are free of the detection loophole, as no postselection is performed.

\begin{figure} [!t]
\centering
\includegraphics[width=0.98\linewidth]{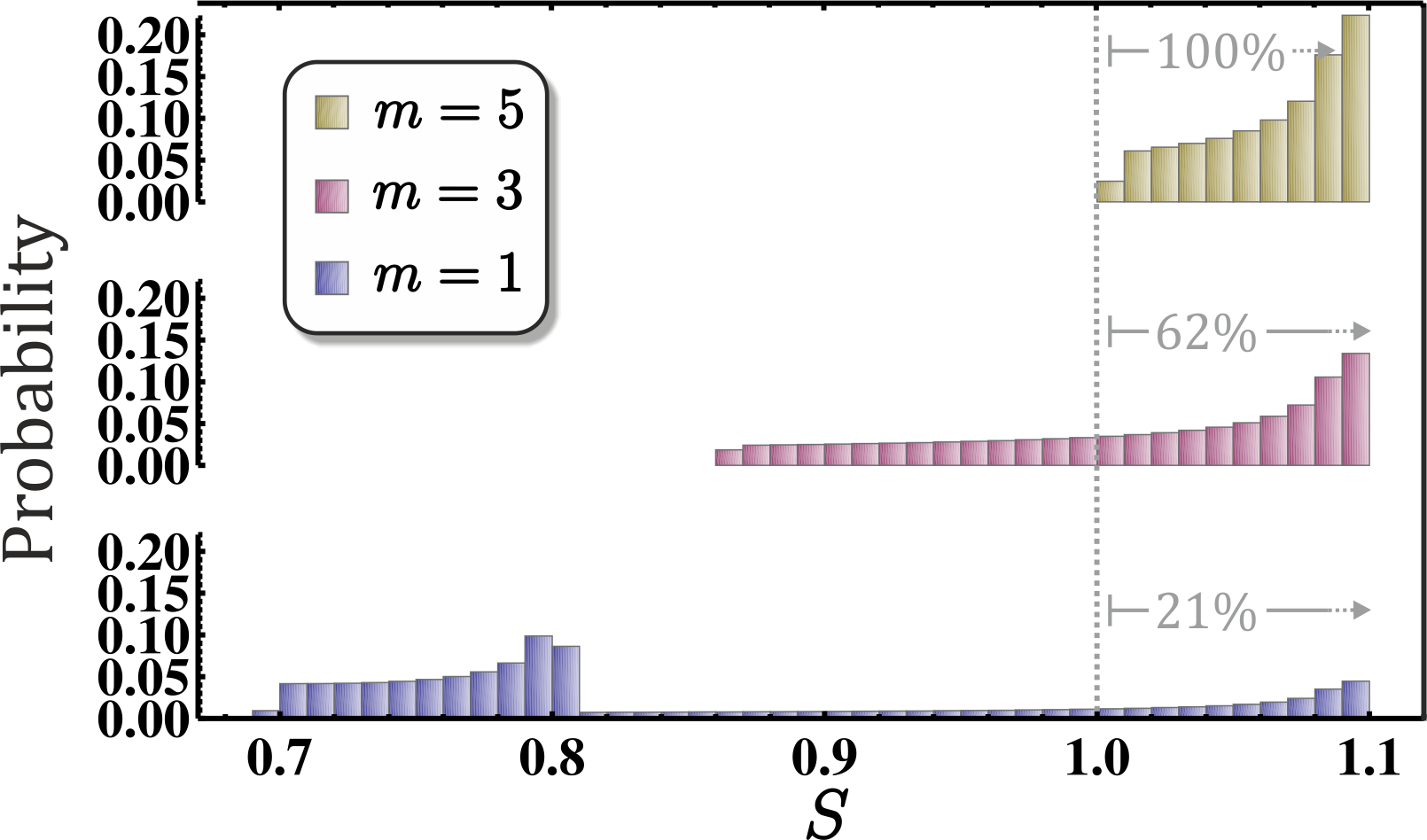}
\caption {(Color online) Distribution of the Bell value $S$ for two parties sharing no common reference frame, and including phase fluctuations ($\delta=0.4$) and losses ($\eta=0.9$) for $m=1,3,5$ (bottom to top). The first party uses $2m$ measurements and the second uses just 2. For $m=5$, Bell violations ($S>1$) are obtained with certainty. Note that the largest violations are in fact most likely to occur.}
\label{fig.violdistr}
\end{figure}

Both $S_{max}$ and $\tilde{\eta}$ can be obtained only for specific values for the average relative phases $\bar{\phi}_l$. Nevertheless, in the case where the parties share no common reference frame (i.e.~all $\bar{\phi}_l$ are unknown), it is possible to obtain values arbitrarily close to $S_{max}$ and $\tilde{\eta}$ by adding enough measurement settings. More importantly, by adding only few measurements, it is possible to obtain significant Bell violations, even in the presence of losses and phase fluctuations. This is shown in \figref{fig.violdistr}, where we plot the distribution of the Bell value $S$ for $N=2$, considering $\bar{\phi}$ to be uniformly distributed in $\left[0,2\pi\right]$. The parties use the amplitude settings (i.e.~the values of $r$, $r'$) which maximise the violation for $\bar{\phi}=0$. The first party uses $m$ pairs of settings. Each pair of settings uses the same two amplitudes, but a shifted value of the phase---i.e. for pair $j$, the local phase is $\varphi_1^j = j 2\pi /m $. As shown in \figref{fig.violdistr}, for $m=5$ (i.e.~10 settings for the first party and 2 settings for the second), we obtain Bell violation with certainty, considering phase fluctuations ($\delta=0.4$) and losses ($\eta=0.9$). With sufficient local settings, for $\eta=0.9$ certain violation can be attained with fluctuations up to $\delta \approx 0.7$. These values seem within experimental reach. Concerning the creation and detection of single photons, a recent experiment reported total efficiencies of $83 \%$ (with foreseeable improvements) for $\sim 5000$ counts per second \cite{Ramelow2012}. Concerning the stabilization of the local phase, commercially available systems (see e.g. \cite{Alnis2008}) can achieve coherence times of several seconds. Overall, perspectives for a demonstration of the nonlocality of a single photon without the need of common reference frame are promising.

\section{Conclusions}

We have presented a scheme for demonstrating unambiguously that a single photon can exhibit quantum nonlocality. Since our results are robust to the combined effect of phase fluctuations and losses, we believe that experimental perspectives are promising.

In fact, the robustness to our scheme to phase fluctuations deserves more comments. A series of recent works \cite{Liang2010,Shadbolt2012,Wallman2012}, explored the robustness of nonlocality (considering multiparticle entanglement) without a common reference frame, and the practical feasability of such schemes has been experimentally demonstrated \cite{Shadbolt2012,Palsson2012} (see also \cite{DAmbrosio2012}). A comparison with the results of \cite{Shadbolt2012,Wallman2012} highlights the robustness of our scheme. We have checked numerically that the former schemes are not robust to arbitrary Gaussian fluctuations of the relative phase (no violation is possible for $\delta > 0.68$), whereas our scheme can tolerate arbitrarily strong fluctuations, that is, any $\delta < \infty$.

We also note that, while we have focused on displacement based measurements, the ideas presented here should also be relevant for Bell tests and other quantum communication protocols based on homodyne measurements. We have performed numerical studies showing that the schemes of Refs.~\cite{Cavalcanti2011,Morin2012} can also lead to Bell violations with high probability without a common reference frame, and we hope the present results will motivate further research along these lines.

Finally, it would be interesting to adapt the present ideas to the case of massive particles and/or fermionic systems, where particle statistics and superselection rules should be taken into account \cite{Ashhab2007}.

\emph{Acknowledgements.---}%
We thank B.~C. Sanders, B. Sanguinetti, and C. Simon for discussions.
J.~B.~B.~was supported by the ERC starting grant PERCENT. R.C. was supported by the Excellence Initiative of the German Federal and State Governments (grant ZUK 43). N.~B.~was supported by the UK EPSRC, the EU DIQIP, and the Swiss National Science Foundation (grant PP00P2-138917).


%

\end{document}